\documentstyle[aps,prb,epsfig,multicol]{revtex}
\begin{document}
\title{Photoemission studies of Ga$_{1-x}$Mn$_{x}$As:\\ Mn-concentration
dependent properties} \author{H. \AA sklund, L. Ilver, and J. Kanski}
\address{Department of Experimental Physics, Chalmers University of
Technology and G\"{o}teborg University, SE-412 96 G\"oteborg, Sweden}
\author{J. Sadowski} \address{Department of Experimental Physics,
Chalmers University of Technology and G\"{o}teborg University, SE-412
96 G\"oteborg, Sweden\\
and Institute of Physics, Polish Academy of Sciences, PL-02-668
Warszawa, Poland}
\author{R. Mathieu} \address{Department of Materials Science, Uppsala University, Box 534, SE-751 21 Uppsala, Sweden}

\date{Submitted: May 4, 2001} \maketitle

\begin{abstract}
Using angle-resolved photoemission, we have investigated the
development of the electronic structure and the Fermi level pinnning
in Ga$_{1-x}$Mn$_{x}$As with Mn concentrations in the range 1--6\%.  We find that
the Mn-induced changes in the valence-band spectra depend strongly on
the Mn concentration, suggesting that the interaction between the Mn
ions is more complex than assumed in earlier studies.  The relative
position of the Fermi level is also found to be
concentration-dependent.  In particular we find that for
concentrations around 3.5--5\% it is located very close to the
valence-band maximum, which is in the range where metallic
conductivity has been reported in earlier studies.  For concentration
outside this range, larger as well as smaller, the Fermi level is
found to be pinned at about 0.15 eV higher energy.
\end{abstract}


\begin{multicols}{2}

\section{Introduction}

The possibility to include magnetic impurities at relatively high
concentrations in GaAs by means of low-temperature molecular beam
epitaxy (LT-MBE) has opened new exciting prospects of combining
magnetic phenomena with high-speed electronics and optoelectronics. 
The numerous investigations of Ga$_{1-x}$Mn$_{x}$As alloys that have
been carried out so far have revealed interesting material properties,
the most notable being carrier-induced ferromagnetism, with reported
Curie temperatures as high as 110 K.\cite{Ohno01} Other interesting
properties are anomalous Hall effect, negative magnetoresistance and
photoinduced ferromagnetism.  Although there is general consensus
concerning the importance of Mn-induced holes, the detailed mechanisms
behind the ferromagnetic ordering of the Mn spins remain a subject of
debate.\cite{Dietl01,Inoue00,Konig00,Akai98,Litvinov01,Schneider87}
The electronic state of the Mn ions in samples with high Mn content is
also discussed, though at low concentrations (below 1\%) the
$d^{5}$+hole configuration is established.\cite{Sanvito00,Szczytko99}
It is clear that further spectroscopic studies related to these
problems are strongly motivated.  In the present work we have used
photoemission to study two key features in the electronic structure of
Ga$_{1-x}$Mn$_{x}$As alloys for a range of Mn concentrations: the
Mn-related modifications of the electronic structure and the Fermi
level position relative to the VBM.

\section{Experiment}

The experiments were performed on the toroidal grating monochromator
beamline (BL 41) at the MAX I storage ring of the Swedish National
Synchrotron Radiation Center MAX-lab, where a dedicated system for
molecular beam epitaxy (MBE) is attached to the photoelectron
spectrometer.  This configuration allows samples to be transferred
between the growth and analytical chambers under UHV conditions.  In
the transfer system the vacuum was in the low 10$^{-9}$ torr range,
and in the electron spectrometer in the low 10$^{-10}$ torr range. 
The ability to transfer samples means that no post-growth treatment
was needed to prepare the surfaces for the photoemission measurements. 
This is a point worth stressing in the present context, as the samples
are prepared under rather extreme conditions and change their
properties with annealing even at temperatures well below that at
which MnAs segregates.\cite{Hayashi01} Indeed, the spectra presented here are
somewhat different from those obtained on sputtered and annealed
surfaces.\cite{Okabayashi99}

The MBE system contains six sources, including an As$_{2}$ valved
cracker.  It is also equipped with a 10 keV electron gun for
reflection high-energy electron diffraction (RHEED).  The samples were
approximately 10$\times$10 mm$^{2}$ pieces of epi-ready $n$-type
GaAs(100) wafers, which were In-glued on tranferrable Mo holders. 
Each sample preparation started with a 1000 {\AA} buffer, grown at a
substrate temperature ($T_{s}$) of 590 $^{\circ}$C, $T_{s}$ was then
lowered to the growth temperature of LT-GaAs and GaMnAs, which was
typically 220 $^{\circ}$C. At the low temperature the growth started
always with a 200--300 {\AA} LT-GaAs buffer layer.  The As$_{2}$/Ga
flux ratio was maintained at values around 10.  During deposition of
this layer the LT-GaAs growth rate was measured by recording RHEED
intensity oscillations.  After opening the Mn shutter the RHEED
oscillations were observed again during the GaMnAs
growth.\cite{Sadowski00} At this low growth temperature the
reevaporation of Mn and Ga from the surface is negligible, so the
growth rate increase is proportional to the Mn content.  The
Mn-concentrations quoted below are estimated to be accurate within
0.5\%.

Immediately after transfer the surfaces were checked with low energy
electron diffraction (LEED).  All Mn-containing samples exhibited
1$\times$2 reconstructed surfaces in RHEED as well in LEED, while the
clean reference GaAs sample displayed a $c$(4$\times$4) LEED pattern
with sharp integer order and less distinct fractional order spots. 
Photoemission was excited with mainly $p$-polarized light incident at
45$^{\circ}$ relative to the surface normal, the samples being
oriented with the [110] azimuth (i.e. the 1-fold periodicity) in the
plane of incidence.  The electron energy distribution curves were
obtained using a hemispherical electron energy analyzer with an
angular resolution of 2$^{\circ}$, and the overall energy resolution
was around 0.3 eV. A clean Ta foil, in contact with the sample holder,
was used to determine the Fermi level position in each case.  The
counting rates were normalized to the incident beam intensity by means
of photocurrent from a gold mesh in the beam path.

\section{Results and discussion}

\begin{figure}[h]
\centerline{\epsfig{figure=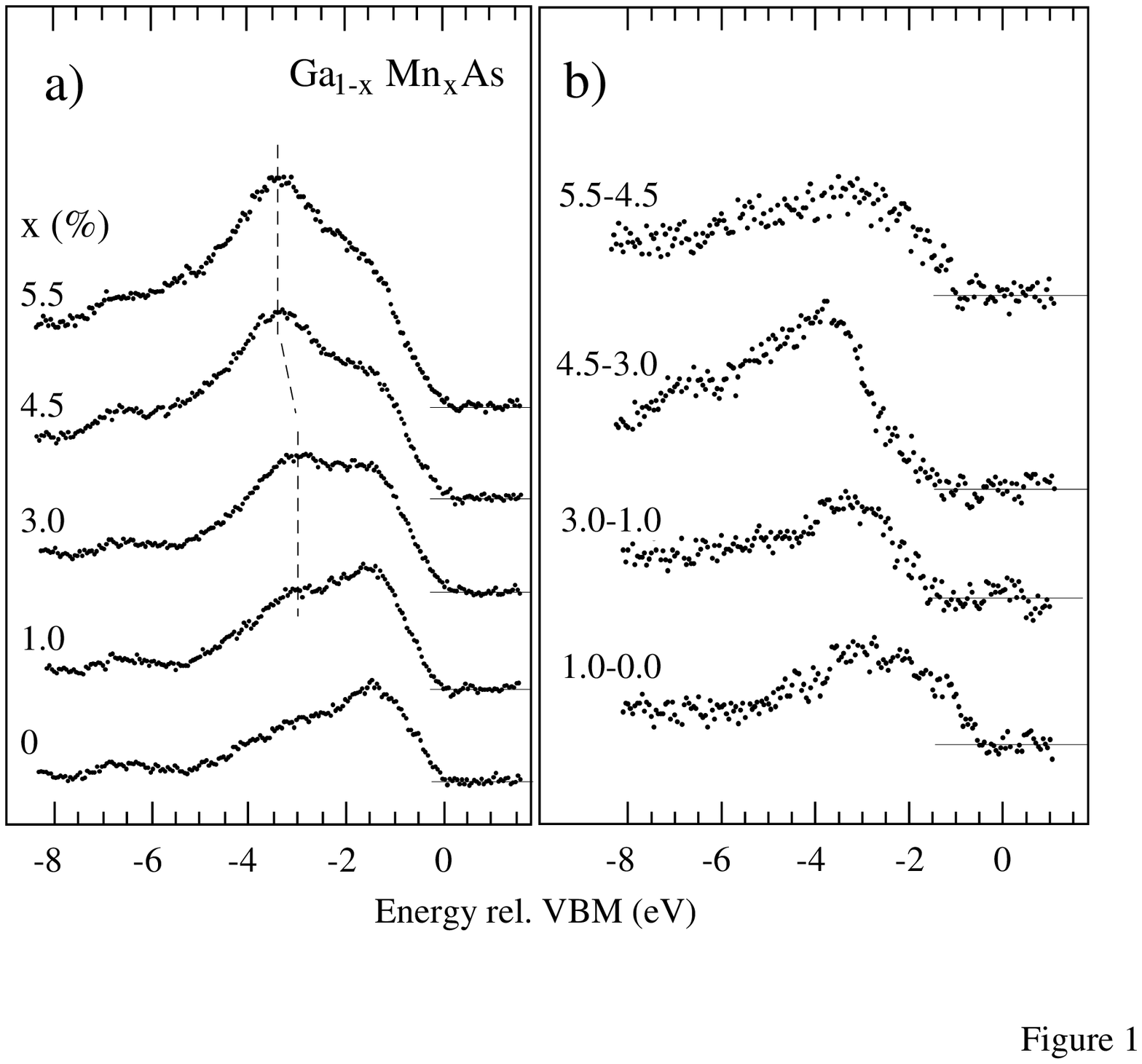,height=14cm,width=12cm,angle=90}}
\caption{(a) Normal emission valence-band photoemission from Ga$_{1-x}$Mn$_{x}$As with different
Mn concentrations, aligned at the valence band maximum.  The spectra
were excited with 81 eV photons.  (b) Difference spectra obtained by
subtracting consecutive spectra in (a).}
\label{Mn}
\end{figure}

Considering the intrinsic surface sensitivity of photoemission, and
that surface compositions in alloy systems often deviate from those in
the bulk, it is well motivated to start with a brief comment on this
point.  The fact that clear and actually unusually persistent RHEED
oscillations are observed during growth of GaMnAs, shows that the
atoms are still mobile in the surface layer despite the
low-temperature conditions.  However, once accommodated in lattice
sites, further mobility that would lead to phase separation is
efficiently inhibited under the low-temperature growth conditions. 
Thus, it is well motivated to expect that the sample compositions are
uniform, including the first atomic layers.  We should mention,
however, that applying secondary ion mass spectroscopy (SIMS) as well
as Auger microprobe analysis on samples exposed to atmospheric
pressure, we have observed pronounced enrichment (by a factor of 2) of
Mn in the surface layer (and a corresponding depletion in the
underlying region), which is clearly associated with oxidation. 
Typically these redistributions range over a thickness of around 150
{\AA}.  This clearly emphasises the significance of carrying out
surface sensitive experiments on {\it in situ} prepared samples. 
Photoemission from Mn 3$d$ states in dilute systems like
Ga$_{1-x}$Mn$_{x}$As is easily identified via resonant enhancement of
the 3$d$ cross section at the 4$p$ excitation threshold, which occurs
at 50 eV photon energy.  Since the spectral shape changes quite much
in this energy range, and since our aim is to compare spectra recorded
from a series of different samples, we have chosen to use a photon
energy well above this resonant range (81 eV).  Although the absolute
cross section of Mn 3$d$ is smaller at 81 eV than that just above 50
eV photon energy, the cross sections of the GaAs valence states are
also reduced in a similar way and therefore the Mn 3$d$-induced
spectral features are still readily detected.  Fig.\ \ref{Mn}a shows a
set of such valence band spectra from samples with different Mn
contents, together with a reference spectrum from clean LT-GaAs.  It
is worth pointing out that just like the GaMnAs, such LT-GaAs contains
large concentrations of point defects (mostly As antisites), and that
spectra from such layers are found to be somewhat different relative
those obtained from MBE layers grown at high
temperature.\cite{Aasklund01} Considering that until now only one
independent valence-band photoemission study of such materials has
been published,\cite{Okabayashi99} and that the samples are produced
under rather extreme growth conditions, it is well motivated to start
the discussion with a direct comparison between the present data and
the published ones.  We note then that the data contain some
similarities, but also some significant differences.  The main
Mn-induced feature is the peak at 3.4 eV below the VBM for the most
Mn-rich samples.  Its Mn origin is clearly revealed by the resonant
enhancement mentioned above.  A similar resonant structure was found
in Ref.\ \ref{RefOkabayashi99}, but at a binding energy of 4.5 eV
relative to the Fermi level.  Since the Fermi level is located about
0.13 eV above the VBM (see below), there seems to be a discrepancy of
almost 1 eV between the two results.  Furthermore, the spectra in
Ref.\ \ref{RefOkabayashi99} contain a second pronounced peak at about
2.5 eV larger binding energy.  This structure is completely missing in
our data.  The weak asymmetric peak seen in all spectra around 6.5--7
eV in Fig.\ \ref{Mn}a reflects the $X_{3}$ critical point emission.  Such
density of states (DOS) structures are seen at all photon energies and
all emission angles due to diffuse elastic scattering of the direct
interband excitation of this state.  Altogether we thus find that the
present spectra are significantly different from those found in
literature, and although the reason for these deviations is not clear,
it is natural at this point to suspect that the different surface
preparations could be the cause.  This would then underline the
importance of carrying out these experiments on {\it in situ} grown
samples.

It is immediately clear in Fig.\ \ref{Mn} that the Mn-induced spectral
changes vary with the Mn content.  To examine this variation in some
more detail, we have generated consecutive difference spectra, as
displayed in Fig.\ \ref{Mn}b.  The first spectrum in this sequence shows that
with 1\% Mn the spectral intensity is increased over a range 1--4 eV
below the VBM, with a peak centered around 3 eV and a shoulder at 1 eV
below the VBM. The main increase coincides with weak structures in the
clean GaAs spectrum (around 3 eV and 4 eV below the VBM).  As these
structures are due to excitations at the high DOS regions at the
$X_{5}$ and $\Sigma^{min}_{1}$ points, one could suspect that the
Mn-induced changes are in this case caused by disorder-related
increase of diffuse scattering.  However, from the fact that no
corresponding increase is seen for the $X_{3}$ critical point
emission, and from the following development of the 3-eV peak, we can
safely conclude that these spectral changes do indeed reflect
Mn-derived states.  With the Mn content raised to 3\% we see that the
incremental change is somewhat different than the initial one.  The
peak at 3 eV is increased further, but the range around 1 eV remains
essentially unchanged.  Increasing the Mn content further results in
another change: the main additional spectral contribution appears as
an asymmetric peak around 3.8 eV below the VBM, i.e. clearly shifted
relative to that found at lower concentrations.  Thus, the peak
observed at 3.4 eV in the corresponding full spectrum (Fig.\ \ref{Mn}) can
be concluded to represent an average of several contributions. 
Finally, the additional spectral changes with further increase of Mn
content are found to be less distinct, the intensity is increased
rather uniformly over a range 2--6 eV below the VBM. The next
difference spectrum is also essentially a peak centered at 3 eV below
the VBM, though it is clearly narrower on the high-energy side.

The important conclusion from the data in Fig.\ \ref{Mn} is that the
character of the Mn states in Ga$_{1-x}$Mn$_{x}$As depends on the Mn
concentration.  Since supplementary X-ray diffraction analysis of our
samples shows high degree of perfection in the layers, we have no
reason to suspect that the variations seen here are due to varying
sample structure quality, but ascribe them to the different Mn
contents.  No such dependence has been reported in any of the earlier
studies.  Previous analysis of Mn 3$d$ partial DOS in GaMnAs with
6.9\% Mn was successful in modelling the observed spectrum using a
configuration interaction model involving Mn 3$d$ and ligand states in
a MnAs4 cluster.\cite{Okabayashi99} Obviously, this kind of model can
not account for concentration-dependent properties like those reported
here.  With an average distance between two impurities of around 25
{\AA}, it is clear that the explanation must be based on a model in
which long range interactions are taken into account.

\begin{figure}[h]
\hspace*{1cm}
\centerline{\epsfig{figure=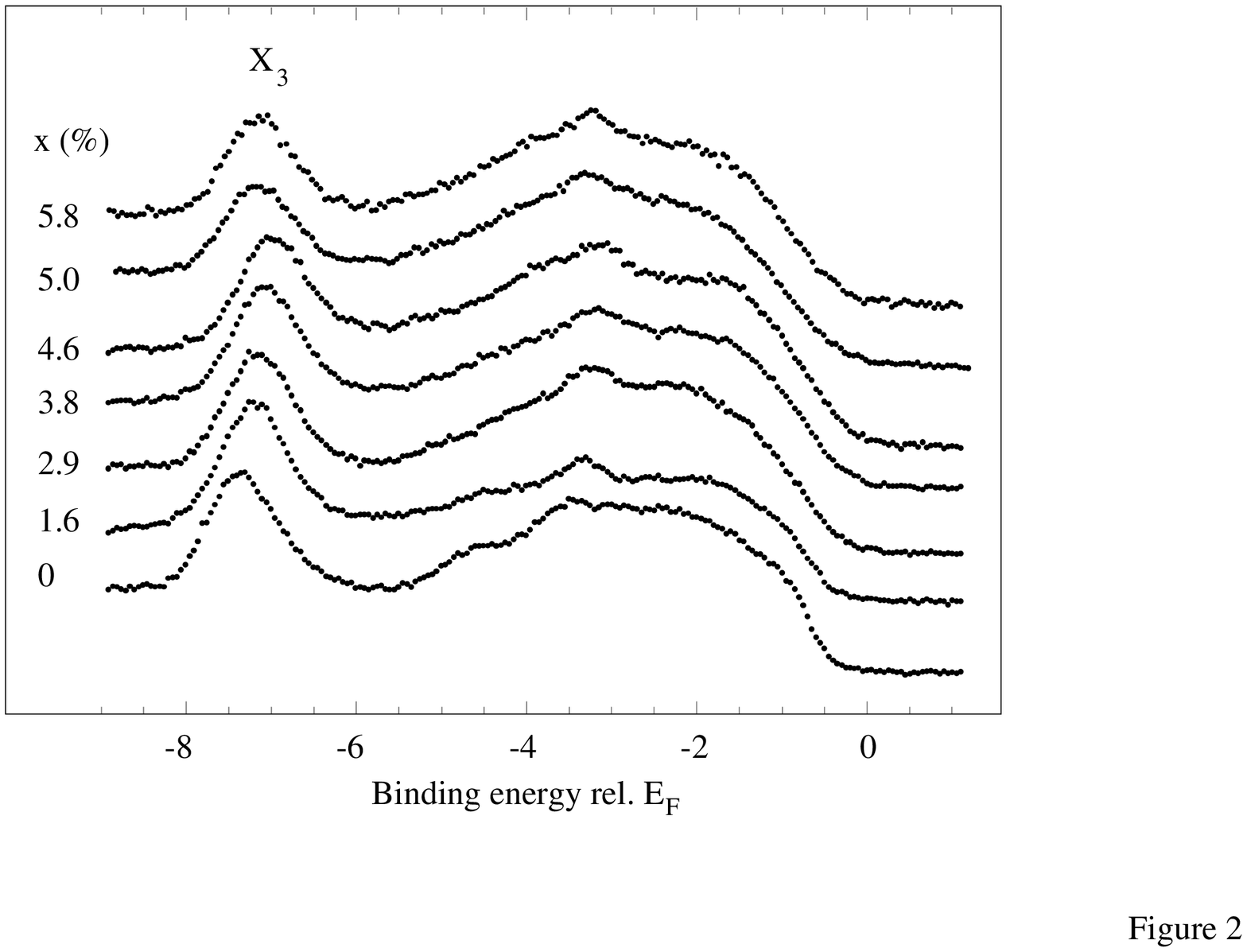,height=14cm,width=12cm,angle=90}}
\caption{Normal emission
valence-band photoemission from Ga$_{1-x}$Mn$_{x}$As with different Mn
concentrations, aligned at the Fermi level.  The spectra were excited
with 38.5 eV photons.}
\label{VB}
\end{figure}

\begin{figure}[h]
\centerline{\epsfig{figure=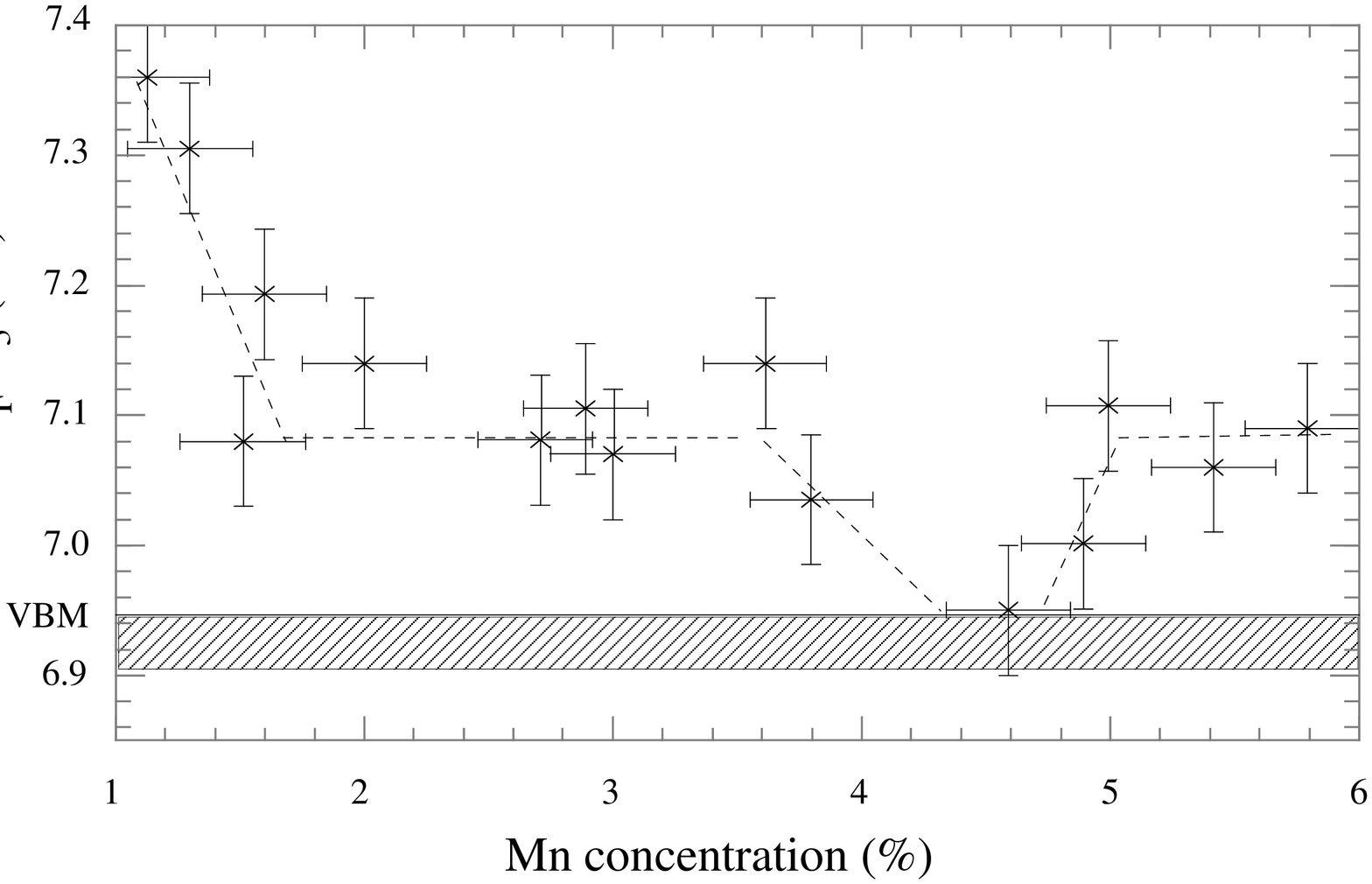,height=14cm,width=10cm,angle=90}}
\caption{Fermi level position relative the $X_{3}$ critical point as a
function of Mn concentration in Ga$_{1-x}$Mn$_{x}$As.  The dashed
region represents the valence band edge.}
\label{EF}
\end{figure}

GaMnAs is also known to exhibit unusual conductivity
characteristics:\cite{Oiwa97} at low Mn concentrations the system is
semiconducting, around 4--5\% Mn metallic conductivity is reported and
with further increase of the Mn content the material becomes again
insulating.  Interestingly enough, the Curie temperature also exhibits
a maximum around the same Mn concentration.  These two observations
suggest that the density of holes is actually decreasing with Mn
concentrations above 5\%, and this might be directly reflected by the
Fermi level position relative the VBM. In Fig.\ \ref{VB} we show a set
of valence-band spectra from samples with varying Mn contents, aligned
at the Fermi level.  The photon energy used in this case was 38.5 eV,
chosen to probe the phase space region around the $X_{3}$ point.  This
emission is reflected by the prominent peak around 7.5 eV. Considering
the high density of defects in LT-GaAs and in GaMnAs, (in the range of
10$^{20}$/cm$^{2}$), it is reasonable to assume that the surface Fermi
level does not deviate from that in the bulk.  This assumption is
supported by the fact that no additional spectral broadening that
could be expected due to a emission from a very narrow band bending
region was detected in any spectra.  Focusing on the $X_{3}$ emission
we see that its position is changing with Mn content.  This variation
is shown more clearly in Fig.\ \ref{EF}, where we have plotted the
energy separation $E_{\text{F}}-X_{3}$ for a larger set of samples. 
Starting at a value of 7.35 eV for clean LT-GaAs it is reduced, and
settles at a value of 7.1 eV around 1.5\% Mn concentration.  This
pinning position remains stable for Mn concentrations up to around
3.5\%, and is likely due to the Mn acceptor level known to be located
113 meV above the VBM.\cite{Schneider87} With this interpretation we
deduce the VBM to be located around 6.95 eV above the $X_{3}$ point, a
value well in the range of literature data\cite{Chiang80} (6.70--7.1
eV) based on angle resolved photoemission and X-ray photoemission.  A
very interesting feature is observed around 4--5\% Mn concentrations,
where $E_{\text{F}}$ appears to drop to a position close to the VBM.
As already mentioned, samples in this concentration range are reported
to exhibit metallic conductivity.  The present observations are fully
consistent with such behaviour.  We also note that the low position of
$E_{\text{F}}$ implies an increased density of holes, which in turn
may be the explanation for the relatively high Curie temperatures
found in this range of Mn concentrations.  The most intriguing
observation is the shift of $E_{\text{F}}$ back into the band gap
region with further increased Mn content.  This is consistent with the
reported metal-insulator transition,\cite{Oiwa97} though the present
results suggest that the reason for the insulating properties is not
impurity scattering, but rather a true reduction of charge carriers.

\section{Conclusions}

The present investigations of valence band photoemission from
Ga$_{1-x}$Mn$_{x}$As compounds show two new effects.  Firstly we find
that the spectral changes induced by the Mn atoms depend on the Mn
concentration, and secondly we observe that the position of the Fermi
level also changes with Mn content.  None of these features has been
reported previously.  The varying shape of the Mn-induced valence-band
structures directly shows that the Mn-host interaction cannot be
treated with a local model.  As to the Fermi level variations, we note
that the minimum observed around 3.5--5\% Mn content coincides with
previously reported metallic conductivity and also with the range of
maximum paramagnetic-ferromagnetic transition temperatures.

\section*{Acknowledgements}

We are pleased to acknowledge the technical support of the MAX-lab
staff.  This work was supportet by grants from the Swedish Natural
Science Research Council (NFR), the Swedish Research Council for
Engineering Sciences (TFR), and, via co-operation with the Nanometer
Structure Consortium in Lund, the Swedish Foundation for Strategic
Research (SSF).

\end{multicols}

\newpage

\end{document}